\documentclass[11pt]{article}
\pdfoutput=1 
\usepackage{mathtools}
\usepackage{booktabs}
\usepackage{amsmath,amssymb,amsbsy,amstext, amsthm, simplewick, amsfonts,braket}
\usepackage{graphicx}
\usepackage[small]{caption}
\usepackage{siunitx}
\usepackage{subcaption}
\usepackage{upgreek}
\usepackage{framed}
\usepackage{wrapfig}
\usepackage{multirow}
\usepackage{bbm}
\usepackage[numbers,sort&compress]{natbib}
\usepackage[svgnames,dvipsnames,x11names]{xcolor}
\usepackage[utf8x]{inputenc}
\usepackage{selinput} 
\usepackage{bm}
\usepackage{float}
\usepackage{dsfont}
\usepackage[margin=1cm]{caption}
\usepackage{subcaption}
\usepackage{sidecap}
\usepackage{longtable}
\usepackage{anyfontsize}

\usepackage{epstopdf}
\usepackage{cancel}
\usepackage{tcolorbox}
\usepackage{latexsym,amsmath,amssymb,epsfig}
\usepackage{braket}
\usepackage{tensor}
\usepackage{tocloft}
\usepackage[permil]{overpic}

\usepackage{tcolorbox}
\definecolor{greyish2}{rgb}{.96,.96,.96}

\def\xyma{\xymatrix@M.7em}
\def\xymas{\xymatrix@M.1em}

\newcommand{\Comment}[1]{{}}
\definecolor{darkblue}{rgb}{0.15,0.35,0.55}
\definecolor{reddish}{rgb}{0.65, 0.2, 0.2}
\definecolor{darkgreen}{RGB}{50,150,0}
\definecolor{greyish2}{rgb}{.96,.96,.96}
\usepackage[linktocpage=true]{hyperref}
\hypersetup{
colorlinks=true,
citecolor=darkblue,
linkcolor=reddish,
urlcolor=darkblue,
pdfauthor={},
pdftitle={},
pdfsubject={}
}

\DeclareFontFamily{OT1}{rsfs10}{}
\DeclareFontShape{OT1}{rsfs10}{m}{n}{ <-> rsfs10 }{}
\DeclareMathAlphabet{\mathscript}{OT1}{rsfs10}{m}{n}

\newcommand\blfootnote[1]{%
  \begingroup
  \renewcommand\thefootnote{}\footnote{#1}%
  \addtocounter{footnote}{-1}%
  \endgroup
}

\def\gsim{ \lower .75ex \hbox{$\sim$} \llap{\raise .27ex \hbox{$>$}} }
\def\lsim{ \lower .75ex \hbox{$\sim$} \llap{\raise .27ex \hbox{$<$}} }
\def\be{\begin{equation}}
\def\ee{\end{equation}}
\def\bea{\begin{eqnarray}}
\def\eea{\end{eqnarray}}

\def\K{K}
\newcommand{\baaa}{\begin{eqnarray}}
\newcommand{\eaaa}{\end{eqnarray}}

\newcommand{\T}{\mathcal{T}}

\DeclareMathOperator{\E}{e}

\usepackage{tikz}
\usetikzlibrary{decorations}
\pgfdeclaredecoration{complete sines}{initial}
{
    \state{initial}[
        width=+0pt,
        next state=upsine,
        persistent precomputation={\pgfmathsetmacro\matchinglength{
            \pgfdecoratedinputsegmentlength / int(\pgfdecoratedinputsegmentlength/\pgfdecorationsegmentlength)}
            \setlength{\pgfdecorationsegmentlength}{\matchinglength pt}
        }] {}
    \state{upsine}[width=\pgfdecorationsegmentlength,next state=downsine]{
        \pgfpathsine{\pgfpoint{0.25\pgfdecorationsegmentlength}{0.5\pgfdecorationsegmentamplitude}}
        \pgfpathcosine{\pgfpoint{0.25\pgfdecorationsegmentlength}{-0.5\pgfdecorationsegmentamplitude}}
    }
    \state{downsine}[width=\pgfdecorationsegmentlength,next state=upsine]{
        \pgfpathsine{\pgfpoint{0.25\pgfdecorationsegmentlength}{-0.5\pgfdecorationsegmentamplitude}}
        \pgfpathcosine{\pgfpoint{0.25\pgfdecorationsegmentlength}{0.5\pgfdecorationsegmentamplitude}}
}
    \state{final}{}
}

\definecolor{greyish}{rgb}{.90,.90,.90}
\definecolor{greyish2}{rgb}{.96,.96,.96}
\usepackage{xcolor,colortbl}
\usepackage{tcolorbox}

\usepackage[all]{xy}

\def \lll {\langle\!\langle}
\def \rr{\rangle\!\rangle}

\newcommand{\p}{\partial}
\renewcommand{\O}{\mathcal{O}}
\renewcommand{\E}{\mathcal{E}}

\DeclareSymbolFont{matha}{OML}{txmi}{m}{it}
\DeclareMathSymbol{v}{\mathord}{matha}{118}

\newcommand{\del}{\nabla}
\renewcommand{\O}{{\mathcal O}}

\newcommand{\R}{{\cal R}}


\numberwithin{equation}{section}
\begin{document}
%
\renewcommand{\thefootnote}{\fnsymbol{footnote}}
\vspace{0truecm}
\thispagestyle{empty}

\begin{center}
{
\bf\LARGE
Light-ray sum rules and the $c$-anomaly\\
}
%
\end{center}


\begin{center}
{\fontsize{12.7}{18}\selectfont
Thomas Hartman${}^{\rm a}$\blfootnote{\texttt{\href{mailto:hartman@cornell.edu}{hartman@cornell.edu}}}
and Gr\'egoire Mathys${}^{\rm b}$\blfootnote{\texttt{\href{mailto:gregoire.mathys@cornell.edu}{gregoire.mathys@epfl.ch}}} 
}

%

\vspace{.8truecm}

\centerline{{\it ${}^{\rm a}$Department of Physics,}}
 \centerline{{\it Cornell University, Ithaca, NY 14850, USA} } 
 
\vspace{.4cm}

\centerline{{\it ${}^{\rm b}$Fields and Strings Laboratory, Institute of Physics}}
 \centerline{{\it Ecole Polytechnique Fédéral de Lausanne (EPFL)} } 
  \centerline{{\it CH-1015 Lausanne, Switzerland} } 

 \vspace{.25cm}

\vspace{.3cm}

\end{center}

\vspace{0.7cm}

\begin{abstract}
\noindent
In a four-dimensional quantum field theory that flows between two fixed points under the renormalization group, the change in the conformal anomaly $\Delta a$ has been related to the average null energy. We extend this result to derive a sum rule for the other anomaly coefficient, $\Delta c$, in terms of the stress tensor three-point function. While the sum rule for $\Delta a$ is an expectation value of the averaged null energy operator, and therefore positive, the result for $\Delta c$ involves the off-diagonal matrix elements, so it does not have a fixed sign.

\end{abstract}

\newpage

\setcounter{page}{2}
\setcounter{tocdepth}{2}
\tableofcontents
\renewcommand*{\thefootnote}{\arabic{footnote}}
\setcounter{footnote}{0}

\section{Introduction}

The trace anomaly in four dimensions has two independent coefficients, $a$ and $c$, known as the type-A and type-B anomalies, respectively \cite{Deser:1976yx,Duff:1977ay}. The $a$ anomaly is monotonic under renormalization group flow \cite{Cardy:1988cwa, Osborn:1989td, Komargodski:2011vj}, famously extending Zamolodchikov's $c$-theorem \cite{Zamolodchikov:1986gt} to four dimensions. In previous work \cite{Hartman:2023qdn}, we showed that the four-dimensional $a$-theorem can be derived from the averaged null energy condition (ANEC) in the form $\Delta a = \langle \Theta | \int du T_{uu} | \Theta\rangle$ in a specially designed state $|\Theta\rangle$ created by a smeared insertion of the trace operator $\Theta$ of the quantum field theory. Here $u$ is a null coordinate. The two-dimensional $c$-theorem can be re-derived from the ANEC in a similar way, along with an infinite set of inequalities on the derivatives of the $c$-function \cite{Hartman:2023ccw}.

The change in the type-B anomaly, $\Delta c$, is not monotonic under the renormalization group (there are known counterexamples \cite{Anselmi:1997ys}), but it nonetheless harbors universal data about the quantum field theory. In this paper, we apply the methods developed in \cite{Hartman:2023qdn} to derive a sum rule relating $\Delta c$ to the three-point function $\langle \Theta \int du T_{uu} T_{ij}\rangle$, where $i$ and $j$ label spatial directions that are transverse to the light ray. See equations \eqref{csumrule}-\eqref{asumruleBB}. This involves an off-diagonal matrix element of the averaged null energy operator, so it does not have a fixed sign, but it suggests that $\Delta a$ and $\Delta c$ are not independent.

Unlike $a$, the $c$ anomaly appears in the stress-tensor two point function at separated points. This has been used to construct sum rules for $c$ and $\Delta c$ from the stress tensor two-point function \cite{Zee:1980sj,Adler:1982ri,Cappelli:1990yc,Anselmi:2001yp,Karateev:2020axc}. The motivation to derive another sum rule for $\Delta c$ that uses the three-point function is twofold. First, it connects to the derivation of the $a$-theorem in \cite{Hartman:2023qdn} and the broader study of light-ray operators, which have sparked a lot of recent interest \cite{Hofman:2008ar,Faulkner:2016mzt,Casini:2017roe,Meltzer:2018tnm,Kravchuk:2018htv,Kologlu:2019bco,Belin:2019mnx,Besken:2020snx,Lee:2022ige,Csaki:2024joe,Chen:2024iuv}. Second, it is interesting to compare light-ray methods to the $S$-matrix. The results we find are analogous to the $S$-matrix sum rules derived for $\Delta a$ in \cite{Komargodski:2011vj} and $\Delta(a-c)$ in \cite{Karateev:2023mrb} by coupling the QFT to a background dilaton and a background metric. The $S$-matrix sum rule for $\Delta a$ is positive by the optical theorem, while $\Delta(a-c)$ is sign-indefinite as it relates to off-diagonal matrix elements. The $S$-matrix methods are based on stress tensor four-point functions, while our approach uses only the stress tensor three-point functions. As we have emphasized in \cite{Hartman:2023qdn}, it is nontrivial that they agree, and interesting to note that this relates the matrix elements of light-ray operators to forward amplitudes in a way that is similar to the Lorentzian inversion formula in CFT \cite{Caron-Huot:2017vep,Hartman:2016lgu,Kravchuk:2018htv}.

Our approach is conceptually a straightforward extension of \cite{Hartman:2023qdn} to include additional matrix elements of the averaged null energy operator. We will also expand on some technical points in the derivation of the sum rule for $\Delta a$, and present a unified derivation of both sum rules simultaneously. In section \ref{s:cft} we consider a conformal fixed point and show how the anomaly coefficients $a$ and $c$ can be extracted from the contact terms in the three-point function $\langle \Theta T_{\mu\nu} T_{\alpha\beta}\rangle$. In section \ref{s:qft} we study this three-point function in a non-conformal QFT and discuss to what extent it is constrained by the CFTs in the UV and IR. In section \ref{s:rg} we derive the sum rule for $\Delta c$. In the final section, we summarize the sum rules obtained in our other recent work on this subject, discuss our sum rule for $\Delta a-\Delta c$ and compare briefly to the $S$-matrix approach used in \cite{Karateev:2023mrb}.

\subsubsection*{Conventions}

In Euclidean signature, we use coordinates 
\be 
ds^2 = d\tau^2 + dy^2 + d\vec{x}^2 = d\tau^2 + dy^2 + \delta_{ij} dx^i dx^j\, ,
\ee
with $\vec{x} \in \mathbb{R}^2$. Greek indices run over all four spacetime dimensions while Latin indices $i,j,{\small\dots} = 1,2$ run over the directions transverse to the $(\tau,y)$-plane. In Lorentzian signature, we use the null coordinates 
\be \label{nullcoord}
ds^2 = -du dv + d\vec{x}^2 = -du dv + \delta_{ij}dx^i dx^j \ ,
\ee
with the Wick rotation $u = t-y = -(y+i\tau)$, $v=t+y = y - i\tau$.
Our convention for the stress tensor is
 $\langle T_{\mu\nu}\rangle = - \frac{2}{\sqrt{g}} \frac{\delta}{\delta g^{\mu\nu}} \log Z$, and we use the conventions of \cite{Hartman:2023qdn} for the metric variations that define correlation functions. Euclidean momenta are denoted by uppercase $K^\alpha$, with the Fourier transform $f(K) = \int d^d x\, e^{iK \cdot x}f(x)$, while Lorentzian momenta are denoted by lowercase $k^\alpha$. Correlators with the momentum-conserving delta function removed are written with double brackets,
\be 
\langle \O(K_1)\dots \O(K_n)\rangle \equiv (2\pi)^d \delta^{(d)}(K_1+\cdots+K_n)\lll \O(K_1)\cdots \O(K_n)\rr \ .
\ee

\section{Anomalies from the averaged null energy in CFT}\label{s:cft}
In this section we study the three-point function $\langle \Theta T_{\mu\nu}T_{\alpha\beta}\rangle$ at a conformal fixed point, first in Euclidean and then in Lorentzian signature, and derive integral formulae to extract the anomaly coefficients $a$ and $c$ from this correlator.

\subsection{Euclidean signature}

In a CFT, the trace of the stress tensor vanishes as an operator equation, $\Theta \equiv T_\mu^\mu = 0$. This means that any correlation function involving at least one trace $\Theta$ can only be supported at coincident points, and is thus at most pure contact terms. The contact terms can be calculated explicitly by varying the conformal anomaly. In four dimensions, in a curved background, the trace is \cite{Deser:1976yx,Duff:1977ay} 
\be \label{theta4dmain}
\langle \Theta\rangle= -a E_4 + c W_{\mu\nu\rho\sigma}^2 + b_1 \Box R + b_2 \Lambda^2 R + b_3 \Lambda^4 \ ,
\ee
where  $\Lambda$ is a UV cutoff, $E_4$ is the Euler density and $W_{\mu\nu\rho\sigma}$ is the Weyl tensor:
\begin{align}
E_4 &= R_{\mu\nu\rho\sigma}^2-4R_{\mu\nu}^2+R^2\, ,\label{eq:Curvature3}\\
W_{\mu\nu\rho\sigma}^2 &= R_{\mu\nu\rho\sigma}^2 -2R_{\mu\nu}^2 +\frac{1}{3}R^2\label{eq:Curvature4}\, ,
\end{align}
with $R_{\mu\nu\rho\sigma}$ the Riemann tensor, $R_{\mu\nu}$ the Ricci tensor and $R$ the Ricci scalar.
The anomaly coefficients $a$ and $c$ are universal, while the $b_i$ terms in \eqref{theta4dmain} can be removed by local counterterms. It is conventional to set the $b_i$ terms to zero in CFT, but in the context of RG flows we must keep track of them because if they are tuned to zero in the UV, they will be generated in the IR.

Varying \eqref{theta4dmain} with respect to the background metric gives all correlation functions of the form $\langle \Theta T_{\mu\nu}T_{\alpha\beta}\dots\rangle$. 
The first variation gives the 2-point function $\braket{T_{\mu\nu}(x_2)\Theta(x_1)}$ whose Fourier transform is \cite{Hartman:2023qdn}
\begin{align}
\lll T_{\mu\nu}(\K)\Theta(-\K)\rr &= 2b_1\K^2 (\K_{\mu}\K_{\nu} -g_{\mu\nu}\K^2)-2b_2\Lambda^2( \K_{\mu}\K_{\nu}- g_{\mu\nu}\K^2)+b_3 \Lambda^4 g_{\mu\nu} \, .\label{eq:Euclidean4dTT2k}
\end{align}
The 3-point function $\langle \Theta T_{\mu\nu} \Theta\rangle$ was calculated in \cite{Hartman:2023qdn} by varying a second time with respect to the Weyl factor. This was sufficient to extract the anomaly coefficient $a$, but for $c$ we will need the other components, $\langle \Theta T_{\mu\nu} T_{\alpha\beta}\rangle$. 
The second variation of \eqref{theta4dmain} with respect to the metric gives the three-point function as
\be 
\braket{\Theta(x_1)T_{\mu\nu}(x_2)T_{\alpha\beta}(x_3)}  =  \frac{4}{\sqrt{g(x_1)}\sqrt{g(x_2)}\sqrt{g(x_3)}}
 \frac{\delta}{\delta g^{\alpha\beta}(x_3)} \frac{\delta}{\delta g^{\mu\nu}(x_2)}
\big[ \sqrt{g(x_1)}\langle\Theta(x_1)\rangle\big]\, .\label{eq:ToVary1AP}
\ee
After a lengthy but straightforward calculation, the result in momentum space is
\begin{align}\label{fullthreept}
\lll \Theta(K_1) T_{\mu\nu}(K_2) T_{\alpha\beta}(K_3) \rr
&= 4(2a-c) \Pi_{\mu\nu\alpha\beta} - \frac{4c}{3} \Pi_{2\mu\nu} \Pi_{3\alpha\beta} +2c (
Q_{\alpha\mu}Q_{\beta\nu} + Q_{\beta\mu} Q_{\alpha\nu} ) \notag \\
&\phantom{=}+ b_1 \mathbb{T}_{\mu\nu\alpha\beta}^{(b_1)}+b_2\Lambda^2  \,\mathbb{T}_{\mu\nu\alpha\beta}^{(b_2)}+b_3 \Lambda^4 \,\mathbb{T}_{\mu\nu\alpha\beta}^{(b_3)}\, .
\end{align}
We have introduced the transverse momentum structures
\begin{align}
\Pi_{2\mu\nu}  &= K_{2\mu} K_{2\nu} - g_{\mu\nu}K_2^2 , & 
\Pi_{3\alpha\beta} &= K_{3\alpha} K_{3\beta} - g_{\alpha\beta} K_3^2\, , \\
Q_{\alpha\mu} &= K_{2\alpha} K_{3\mu} - g_{\alpha\mu} K_2 \cdot K_3 \, ,&
\Pi_{\mu\nu\alpha\beta} &=  \epsilon_{\alpha \rho \sigma (\mu} \epsilon_{\nu) \beta \kappa \lambda} K_2^\rho K_3^\sigma K_2^\kappa K_3^\lambda \ ,\notag
\end{align}
which satisfy
\be 
K_2^\mu \Pi_{2\mu\nu} = K_3^\alpha \Pi_{3\alpha\beta} = 0 , \qquad
K_2^\mu Q_{\alpha\mu} = K_3^\alpha Q_{\alpha\mu} = 0 , \qquad
K_2^\mu \Pi_{\mu\nu\alpha\beta} = K_3^\alpha \Pi_{\mu\nu\alpha\beta} = 0  \ . 
\ee
The first three terms in \eqref{fullthreept}, which contain all of the dependence on $a$ and $c$, are manifestly transverse; they vanish upon contracting with $K_2^\mu$ or $K_3^\alpha$. The tensor structures  corresponding to the counterterms are
\begin{align}
\mathbb{T}_{\mu\nu\alpha\beta}^{(b_3)}&= g_{\mu \nu } g_{\alpha \beta }+ g_{\mu\alpha  } g_{\nu\beta  }   +g_{\mu\beta  }g_{\nu\alpha }   
 \, ,\\
\mathbb{T}_{\mu\nu\alpha\beta}^{(b_2)}&= 2g_{\mu \nu }g_{\alpha \beta } \left( K_2^2 +  K_2\cdot K_3 + K_3^2\right)+2g_{\mu(\alpha}g_{\beta)\nu}\left(2K_2^2 + 5 K_2\cdot K_3 + 2K_3^2 \right)\\
 &  - 2 g_{\mu \nu }\left(2K_{2\alpha}K_{2\beta} + K_{2\alpha}K_{3\beta}+ K_{2\beta}K_{3\alpha} + K_{3\alpha}K_{3\beta}\right)- g_{\mu \alpha }K_{2\beta}K_{3\nu}- g_{\mu\beta } K_{2\alpha}K_{3\nu} \nonumber\\
&- 2
g_{\alpha \beta }\left(K_{2\mu}K_{2\nu} + K_{2\mu}K_{3\nu}+ K_{2\nu}K_{3\mu} + 2 K_{3\mu}K_{3\nu}\right) -g_{\nu\beta } K_{2\alpha}K_{3\mu} 
- g_{\nu\alpha  } 
K_{2\beta}K_{3\mu}\notag\, ,
\end{align}
and
\begin{align}
\mathbb{T}_{\mu\nu\alpha\beta}^{(b_1)} &=-2 g_{\mu\nu}g_{\alpha\beta}\left(K_2^4+ 2K_2\cdot K_3\left(K_2^2  + (K_2\cdot K_3) +  K_3^2 \right)+ K_3^4 \right)\\
&-2g_{\mu(\alpha}g_{\beta)\nu}\Big(2 K_2^4 + 10 (K_2\cdot K_3)^2 + 9 K_2\cdot K_3(K_2^2 + K_3^2)  + 4 K_2^2 K_3^2+ 2 K_3^4 \Big)\nonumber\\
&+ 2g_{\mu\nu}\Big[2(2K_2^2 + 2K_2\cdot K_3 + K_3^2)(K_{2\alpha}K_{2\beta} + K_{2(\alpha}K_{3\beta)} ) + K_3\cdot (K_2+K_3)K_{3\alpha}K_{3\beta} \Big]\nonumber\\
&+ 2g_{\alpha\beta}\Big[2(2 K_3^2+ 2K_2\cdot K_3+K_2^2  )(K_{3\mu}K_{3\nu} + K_{2(\mu}K_{3\nu)}) +K_2\cdot  ( K_3 + K_2)K_{2\mu}K_{2\nu} \Big]\nonumber\\
&+2\Big[g_{\mu(\alpha}K_{2\beta)}K_{3\nu} + g_{\nu(\alpha}K_{2\beta)}K_{3\mu}\Big](K_2 + K_3)^2\nonumber\\
&-4 \left[K_{2\mu}K_{2\nu}\left(K_{2\alpha}K_{2\beta}+K_{2(\alpha}K_{3\beta)} \right)+ K_{3\alpha}K_{3\beta}\left(K_{3\mu}K_{3\nu}+ K_{2(\mu}K_{3\nu)}\right)\right]\notag\, .
\end{align}
The counterterm structures (on the second line in \eqref{fullthreept}) are not transverse, as expected since these terms satisfy an inhomegenous Ward identity. The Ward identity for this 3-point function, derived in appendix A of \cite{Hartman:2023qdn}, is
\begin{align}\label{wardidentity2}
\del^\alpha\langle  \Theta(x_1) T_{\mu\nu}(x_2) T_{\alpha\beta}(x)\rangle &= \langle T_{\mu\nu}(x)\Theta(x_1)\rangle \del_{\beta} \delta^{(d)}(x-x_2)+ \nabla_\nu\big(\braket{T_{\beta\mu}(x)\Theta(x_2)}\delta^{(d)}(x-x_2)\big)\notag\\
&+ \braket{T_{\mu\nu}(x_2)\Theta(x)}\nabla_\beta \delta^{(d)}(x-x_3)+ \nabla_\mu\big(\braket{T_{\beta\nu}(x)\Theta(x_1)}\delta^{(d)}(x-x_2)\big)\, .
\end{align}
In momentum space, this becomes
\begin{align}
K_3^\alpha \lll \Theta(K_1) T_{\mu\nu}(K_2) T_{\alpha\beta}(K_3) \rr &= -K_{2\beta}\lll T_{\mu\nu}(\K_1)\Theta(-\K_1)\rr-K_{1\beta}\lll T_{\mu\nu}(\K_2)\Theta(-\K_2)\rr\nonumber\\
&\phantom{=}+ K_{3\mu}\lll T_{\beta\nu}(\K_1)\Theta(-\K_1)\rr + K_{3\nu}\lll T_{\beta\mu}(\K_1)\Theta(-\K_1)\rr \, . \label{eq:WI3}
\end{align}
This identity is indeed satisfied by the correlators in \eqref{eq:Euclidean4dTT2k} and \eqref{fullthreept}. 

The full expression \eqref{fullthreept} is cumbersome, but we will not need it (except to compare to \cite{Karateev:2023mrb} in the discussion section) and have only included it for completeness. The part that we will use below is obtained by contracting $T_{\mu\nu}(K_2)$ with the null polarization $u^\mu u^\nu$, where $u^\nu$ points in the $u$-direction, setting $u^\alpha K_{2\alpha} = K_{2u} = 0$, and restricting $\alpha,\beta$ to lie in the directions $i,j=1,2$ transverse to the $uv$-plane. In this case, the 3-point function \eqref{fullthreept} reduces to
\begin{align}
 \lll \Theta(K_1) T_{uu}(K_2) T_{ij}(K_3) \rr\big|_{K_{2u}=0}&= K_{3u}^2 \Big[ 8(c-a) (K_{1i}+ K_{3i})(K_{1j}+K_{3j})
-4b_1 K_{3i} K_{3j} \notag\\
&+ 4 \delta_{ij} \left( (2a-c) (K_1+K_3)^2 + b_1(K_1^2+K_3^2)-b_2\Lambda^2\right)\Big] \ .  \label{mainEuc}
\end{align}

\subsection{Lorentzian signature}

Contact terms in time-ordered, anti-time-ordered, retarded, and advanced Lorentzian correlators are identical to those in Euclidean signature, up to numerical prefactors \cite{Hartman:2023qdn}. This follows from the fact that all of these correlators are related by analytic continuation in momenta to the Euclidean correlator, and since contact terms are analytic, the continuations are trivial. For general three-point functions, the numerical factors are 
\begin{align}\label{contactWick}
\begin{split}
\lll  \R[\O_1(k_1); \O_2(k_2)\O_3(k_3)]\rr_{\rm contact}
&=  G_E(k_1, k_2, k_3) \\
\lll  \T[\O_1(k_1)\O_2(k_2)\O_3(k_3)]\rr_{\rm contact}
&= -  G_E(k_1,k_2,k_3)\, , 
\end{split}
\end{align}
where $G_E(K_1,K_2,K_3) = \lll \O_1(K_1) \O_2(K_2) \O_3(K_3)\rr_{\rm contact}$ are the contact terms in the Euclidean correlator, viewed as a function of any two independent momenta.   Since this function is analytic, it can be evaluated at Lorentzian momenta unambiguously, by simply replacing $K_i \to k_i$. In \eqref{contactWick}, $\mathcal{T}$ is time-ordering, and ${\cal R}$ is retarded ordering, defined at separated points by the nested commutators
\be 
\langle \R[ \O_1(x_1); \O_2(x_2) \O_3(x_3)]\rangle
= -\theta( t_1-t_2)\theta(t_2-t_3) \langle [[\O_1(x_1), \O_2(x_2)], \O_3(x_3)]\rangle
+ (2 \leftrightarrow 3) \, .\ \label{eq:Retardedat3pt}
\ee
We will consider ordered correlators of the averaged null energy (ANE) operator,
\be
\E_u(v,\vec{x}) = \int_{-\infty}^{\infty} du \, T_{uu}(u,v,\vec{x}) \ . 
\ee
When $\E_u$ is written with no arguments, it is located on a light-ray through the origin, and this can be written as an integral over momenta as
\be \label{edefk}
\E_u := \E_u(0,\vec{0}) = \int \frac{dk_v\, d^2 \vec{k} }{\pi(2\pi)^2} T_{uu}(k_u = 0, k_v, \vec{k}) \ . 
\ee
 Ordered correlators of the ANE are defined by ordering inside the light-ray integral,
\begin{align}
\begin{split}
\langle {\mathcal{T}}[\mathcal{E}_u(v,\vec{x})\O(x_1)\O(x_2)]\rangle
&= \int_{-\infty}^{\infty} du\, \langle {\cal T}[T_{uu}(u,v,\vec{x}) \O(x_1)\O(x_2) ] \rangle\label{eq:TimeOrderIntegral} \\
\langle {\cal R}[\mathcal{E}_u(v,\vec{x}); \O(x_1)\O(x_2)]\rangle
&= \int_{-\infty}^{\infty} du\, \langle 
\R[T_{uu}(u,v,\vec{x});  \O(x_1)\O(x_2) ] \rangle \ ,
\end{split}
\end{align}
where $\O$ is any operator.
Using the Fourier representation of the ANE in \eqref{edefk}, we have
\begin{align}\label{e3fourier2}
\langle \mathcal{T}[ \E_u\, \O(k_1) \O(k_3) ] \rangle
&= \int \frac{dk_{2v}\,d^2 \vec{k}_2}{\pi(2\pi)^2} \left. \lll \mathcal{T}[T_{uu}(k_2)   \O(k_1)  \O(k_3) ] \rr\right|_{k_{2u} = 0} \notag \\
& \quad \times \tfrac{1}{2} (2\pi)^4 \delta(k_{1u} + k_{3u}) \delta(k_{1v} + k_{2v} + k_{3v}) \delta^{(2)}( \vec{k}_1+\vec{k}_2 + \vec{k}_3)\, ,
\end{align}
and this implies the identity
\begin{equation}\label{e3fourier}
\langle \mathcal{T}[ \E_u\, \O(k_1) \O(k_3) ] \rangle
=  2\pi  \lll \mathcal{T}[ T_{uu}(-k_1-k_3) \O(k_1)   \O(k_3) ] \rr \delta(k_{1u} + k_{3u}) \ . 
\end{equation}
%
An identical formula holds for $\mathcal{T} \to \mathcal{R}$. The additional factor of $\frac{1}{2}$ on right-hand side of \eqref{e3fourier2} comes from the Jacobian in $\delta^{(4)}(p^\alpha) = \frac{1}{2} \delta(p_u) \delta(p_v) \delta^{(2)}(\vec{p})$.

With these general results in hand we now return to the correlation function $\langle \Theta T_{uu} T_{ij}\rangle$, calculated in a CFT in Euclidean signature in the previous section \eqref{mainEuc}. This is a pure contact term, so it is straightforward to obtain the ordered correlators in Lorentzian signature. Combining \eqref{e3fourier}, \eqref{contactWick}, and \eqref{mainEuc} gives the Lorentzian 3-point function of the averaged null energy,
\begin{align}
\langle \mathcal{R}[ \E_u ; \Theta(k_1) T_{ij}(k_3) ] \rangle &=
- \langle \mathcal{T}[ \E_u\, \Theta(k_1) T_{ij}(k_3) ] \rangle\notag \\
&
=2\pi k_{3u}^2 \Big[  8 (c-a)(k_{1i}+k_{3i})(k_{1j}+k_{3j})-4b_1k_{3i}k_{3j}\label{eq:MainResult1}\\
&\quad +4 \delta_{ij}\Big((2a-c)(k_1 + k_3)^2 +b_1(k_1^2+k_3^2) - b_2 \Lambda^2 \Big)\Big] \delta(k_{1u}+k_{3u})\, . \notag 
\end{align}
\subsection{Extracting the anomaly coefficients}
We will now invert \eqref{eq:MainResult1} to extract the anomaly coefficients $a$ and $c$ from the CFT three-point function by acting with momentum derivatives. This is guided by two considerations. First, we need to isolate momentum structures that are unaffected by the counterterms, $b_i$. Second, as we will see in the next section, for the purposes of deriving an RG sum rule we need to look at terms that are unaffected by semi-local contributions to the QFT correlator or improvement terms. It turns out that all of these requirements are met by the momentum tensor structures in \eqref{eq:MainResult1} involving $k_{1i} k_{3j}$ and $\delta_{ij}\vec{k}_1 \cdot \vec{k}_3$, where $\vec{k}$ denotes the transverse momentum in the spatial directions orthogonal to the null ray. It is clear from \eqref{eq:MainResult1} that these structures have no contribution from the counterterms, and we will show below that they are also insensitive to semi-local terms and improvement terms. Thus we extract the anomalies from \eqref{eq:MainResult1} by first integrating over $k_{1u}$ to remove the delta function, then acting with a linear combination of $\p_{k_{3u}}^2 \p_{k_{1}}^i \p_{k_{3}}^j$ and $\p_{k_{3u}}^2 \delta^{ij} \vec{\p}_{k_1} \cdot \vec{\p}_{k_3}$, and finally setting the momenta to zero. 
The linear combination that extracts the $c$ anomaly is
\begin{align}
c &= \frac{1}{128\pi}\left[\left(\partial_{k_1}^i \partial_{k_3}^j + \frac{1}{2}\delta^{ij}\vec{\partial}_{k_1}\cdot \vec{ \partial}_{k_3} \right)\partial_{k_{3u}}^2\int_{-\infty}^{\infty} dk_{1u} \braket{\mathcal{R}\left[\mathcal{E}_u;\Theta(k_1)T_{ij}(k_3)\right]}\right]_{k_1=k_3=0}\\
&= \frac{1}{64}\int d^4x_1 d^4x_3\, u_3^2\, \delta(u_1)\left(x_1^i x_3^j + \frac{1}{2}\delta^{ij}\vec{x}_1\cdot \vec{x}_3\right) \braket{\mathcal{R}\left[\mathcal{E}_u;\Theta(x_1)T_{ij}(x_3)\right]}\, .\label{cftc}
\end{align}
Recall that the components $i,j$ and vectors $\vec{x}_{1,3}$ are in the transverse spatial directions; see \eqref{nullcoord}.
In the second line we moved the $k$-derivatives inside the Fourier transform that defines the momentum-space correlator. The $a$-anomaly is extracted by the linear combination 
\begin{align}
a &= \frac{1}{128\pi}\left[\delta^{ij} \vec{\partial}_{k_1}\cdot \vec{ \partial}_{k_3} \partial_{k_{3u}}^2 \int_{-\infty}^{\infty} dk_{1u}  \braket{\mathcal{R}\left[\mathcal{E}_u;\Theta(k_1)T_{ij}(k_3)\right]} \right]_{k_1=k_3=0}\\
&= \frac{1}{64}\int d^4x_1 d^4x_3\, u_3^2\, \delta(u_1)(\vec{x}_1\cdot \vec{x}_3) \delta^{ij}\braket{\mathcal{R}\left[\mathcal{E}_u;\Theta(x_1)T_{ij}(x_3)\right]}\, .\label{eq:SumRuleaPT}
\end{align}
Similar formulas hold for the time-ordered correlators, with the replacement ${\cal R} \to -{\cal T}$. Since we are currently considering theories that are conformal, the integrands in the position-space expressions in \eqref{cftc} and \eqref{eq:SumRuleaPT} are pure delta functions --- the role of the integrals is simply to extract the coefficient of the relevant part of the contact term.

\section{Properties of the QFT correlator at low momentum}\label{s:qft}

We now consider the correlation function $\langle \Theta T_{\mu\nu} T_{\alpha\beta}\rangle$ in a quantum field theory that flows between two conformal fixed points. 
Denote the Euclidean correlator in the QFT by
\be \label{propG}
G^{\rm QFT}_{\mu\nu\alpha\beta} (K_1,K_3) :=  \lll  \Theta(K_1) T_{\mu\nu}(-K_1-K_3) T_{\alpha\beta}(K_3) \rr_{\rm QFT} \ . 
\ee
%
%
The corresponding 3-point functions in the CFTs at the fixed points are written $G^{\rm UV}_{\mu\nu\alpha\beta}$ and $G^{\rm IR}_{\mu\nu\alpha\beta}$. Each of the CFT correlators is a pure contact term given in equation \eqref{fullthreept}, with coefficients $(a_{IR}, c_{IR}, b_{i,IR})$ in the IR CFT and $(a_{UV}, c_{UV}, b_{i,UV})$ in the UV. We choose counterterms to set $b_{i,UV} = 0$, and drop the `IR' label on $b_{i,IR} \equiv b_i$. 

The goal of this section is to understand exactly how the QFT correlator $G^{\rm QFT}$ is related to $G^{\rm UV}$ and $G^{\rm IR}$ at the fixed points. Naively, the UV CFT controls the contact terms in $G^{\rm QFT}$, and the IR CFT controls the low-momentum limit of $G^{\rm QFT}$. Neither of these statements is entirely true, due to two complications: First, there are semi-local, partial contact (PC) terms in the correlator, with two points coincident and the third point separated. These are contact terms which are not calculable in the UV CFT. Second, there can be improvement terms to the stress tensor in the IR. 

Ultimately we are interested in Lorentzian correlators of the averaged null energy, so although we are working in Euclidean signature in this section, we set $K_{2u} = -K_{1u} - K_{3u} = 0$, and look at the components $G_{uuij}$
where $u$ is the null direction and $i,j=1,2$ are transverse to the $uv$-plane. In the next two subsections, we show that improvement terms and semi-local terms in this correlator take a very special form, with no contribution to the momentum structures $K_{1i} K_{3j}$ or $\delta_{ij} \vec{K}_1 \cdot \vec{K}_3$. This will be enough to ensure that improvement terms and semi-local terms drop out of the sum rules for $\Delta a$ and $\Delta c$ derived in the next section.\footnote{This was by no means guaranteed --- the specific way of extracting $a$ and $c$ in section \ref{s:cft} was chosen in order to make the improvement terms and PC terms drop out. There are other ways to extract $a$ and $c$ in CFT from the 3-point function \eqref{fullthreept}, but that do not produce useful RG sum rules because they are contaminated by PC terms.  For example, from \eqref{fullthreept} one obtains $\left. \lll \Theta(K_1) T_{uu}(K_2) T_{uu}(K_3)\rr\right|_{K_{1u}=0} =\frac{8c}{3} K_{2u}^4$. This can easily be inverted to write a CFT formula for $c$ that involves the averaged trace, $\int du\, \Theta$, instead of the averaged null energy operator. In superrenormalizable theories this leads to a sum rule for $\Delta c$ along RG flows (and the sum rule is satisfied in the free massive scalar) but we were unable to prove that semilocal contributions from marginal operators drop out. After some experimentation we found momentum structures that were insensitive to this ambiguity, but it would be nice to understand at a more conceptual level why some terms are universal. This is probably related to the question of how the null energy sum rules relate to $S$-matrix amplitudes.
}

The analysis here follows  the same strategy as \cite{Hartman:2023qdn}, while keeping track of the additional terms needed for $\Delta c$.\footnote{It turns out that improvement terms and semi-local terms produce the same potential ambiguities in the correlator, even though they are logically distinct.  They were treated all together in \cite{Hartman:2023qdn}.} 

\subsection{Improvement terms}
If a four-dimensional CFT has a dimension-2 scalar operator $\Phi$, then the stress tensor can be shifted by adding a term proportional to $ (\p_\mu \p_\nu - g_{\mu\nu}\p^2) \Phi $ without spoiling conservation. The statement that a QFT flows to a conformal fixed point in the IR means that $T_{\mu\nu}^{\rm QFT}$ flows to the CFT stress tensor $T_{\mu\nu}^{\rm IR}$ up to a possible improvement term,
\begin{align}\label{ishift}
T_{\mu\nu}^{\rm QFT} \to T_{\mu\nu}^{\rm IR}  + \lambda(\p_\mu \p_\nu - g_{\mu\nu}\p^2) \Phi \ . 
\end{align}
That is, in general, correlation functions of $T_{\mu\nu}^{\rm QFT}$ at low momentum (compared to the mass scale $M$ of the RG flow) can be related to the correlators of the CFT by the shift \eqref{ishift}.  If there are no dimension-2 scalars, or no improvements are needed, then $\lambda=0$ and the QFT stress tensor flows to that of the IR CFT directly.

In the correlator of interest, $\langle \Theta T_{uu} T_{ij}\rangle$, there are possible contributions from improvement terms from shifting each of the three stress tensors. In appendix \ref{app:improvement} we expand out the various terms that come from the three shifts, and enumerate the resulting momentum structures using the fact that $\Phi$ has dimension 2 in the IR. 
At $O(K^4)$, with $K_{2u}=0$, only two terms survive, such that the 3-point functions in the QFT and the IR CFT are related by
\be \label{finalImpA}
G^{\rm QFT}_{uu ij} \approx G^{\rm IR}_{uuij} + 
G_{uuij}^{\mbox{\scriptsize improvement}}\, ,
\ee
where
\be \label{finalImpB}
G_{uuij}^{\mbox{\scriptsize improvement}} = c_1 K_1^2 K_{3u}^2 \delta_{ij} + c_2  (K_{3i} K_{3j} - \delta_{ij} K_3^2)K_{3u}^2 \, ,
\ee
with constants $c_1$ and $c_2$.
Crucially, there are no improvement terms with momentum dependence $K_{1i} K_{3j}$ or $\delta_{ij} \vec{K}_1 \cdot \vec{K}_3$. This will ensure that the improvement terms do not affect the sum rule derived in section \ref{s:rg}.

\subsection{Semi-local terms}
Now we consider semi-local `partial contact' terms. Within a given regularization scheme, the QFT 3-point function \eqref{propG} can be split into three terms, classified according to how many points are coincident:
\be \label{gsplit}
G^{\rm QFT}_{\mu\nu\alpha\beta} = G^{\rm UV}_{\mu\nu\alpha\beta} + G^{\rm PC}_{\mu\nu\alpha\beta} + G^{\rm sep}_{\mu\nu\alpha\beta}\, .
\ee
The first term $G^{\rm UV}$ is the part of the correlator in which all three points coincide; it is equal to the contact term in the UV CFT. The second term $G^{\rm PC}$ is the partial contact term, coming from two points coincident and one separated. The third term is the correlator at separated points. Changing the regulator alters how different contributions appear in the three terms, but for a fixed scheme, the split in \eqref{gsplit} is well defined (see e.g.~\cite{Cappelli:1990yc,Freedman:1991tk}).

We need to understand the partial contact terms that can appear in $G_{uuij}^{\rm PC}$ for $K_{2u} = 0$. 
There are three types of PC terms to consider,
\begin{align}
\braket{\Theta(x_1)T_{\mu\nu}(x_2)T_{\alpha\beta}(x_3)}_{\rm PC} &= \braket{ \lbrace \Theta(x_1)T_{\mu\nu}(x_2)\rbrace T_{\alpha\beta}(x_3)}+\braket{  \Theta(x_1)\lbrace T_{\mu\nu}(x_2) T_{\alpha\beta}(x_3)\rbrace}\nonumber\\
&\quad +\braket{ T_{\mu\nu}(x_2) \lbrace \Theta(x_1)T_{\alpha\beta}(x_3)\rbrace}\, ,\label{eq:PossiblePCTpp}
\end{align}
where the braces show which two points are coincident. The last term, which is proportional to $\delta^{(4)}(x_1-x_3)$, drops out when $T_{\mu\nu}(x_2)$ is contracted with a null polarization and integrated to obtain the averaged null energy operator, because the ANE annihilates the vacuum: $\E_u|0\rangle =0$ \cite{Epstein:1965zza}. See \cite{Hartman:2023qdn} for a more detailed discussion. We will therefore ignore this term and focus on the other two.

Partial contact terms come from operator-valued, local terms in the OPE, proportional to delta functions and derivatives of delta functions.  We therefore analyze their contributions by looking at terms in the OPE of the form
\be 
T_{\mu\nu}(x_2)\Theta(x_1) \supset \mathcal{O}^{\sigma_1\dots\sigma_\ell}(x_2)P_{\mu\nu\sigma_1\dots\sigma_\ell}(\partial)\delta^{(4)}(x_2-x_1)\, ,\label{eq:OPE1}
\ee
and
\be 
T_{\mu\nu}(x_2)T_{\alpha\beta}(x_3) \supset \mathcal{O}^{\sigma_1\dots\sigma_\ell}(x_2)U_{\mu\nu\alpha\beta\sigma_1\dots\sigma_\ell}(\partial)\delta^{(4)}(x_3-x_2)\, ,\label{eq:OPE2}  
\ee
where $\mathcal{O}^{\sigma_1\dots\sigma_\ell}$ is a general local operator of spin $\ell$. These operators are not assumed to be primary, and they can be multiplied by $\beta$-functions and positive powers of the mass scale $M$. The differential operators $P_{\mu\nu\sigma_1\dots\sigma_\ell}$ and $U_{\mu\nu\alpha\beta\sigma_1\dots\sigma_\ell}$ are built out of derivatives and the metric. These OPEs are then plugged into the three-point function \eqref{eq:PossiblePCTpp} and expanded at low momentum.

In appendix \ref{app:semilocal} we enumerate all the possible contributions of this type. After accounting for allowed tensor structures and unitarity bounds, there are just eight possibilities for the spin and IR scaling dimension $(\ell, \Delta_{\O})$ of the $\O^{\sigma_1\dots\sigma_\ell}$ operators. The conclusion of that analysis is that for $K_{2u}=0$, and expanding all of the allowed PC terms at low momentum, there are only three possible contributions at $O(K^4)$:
\be \label{pcfinal}
G_{uu ij}^{\rm PC}(K_1, K_3) \approx  \left( c'_1
K_{3i} K_{3j} 
+ c'_2 \delta_{ij} K_3^2  
+  c'_3\delta_{ij}K_1^2  \right) K_{3u}^2 \, .
\ee
This has the same form as the improvement terms in \eqref{finalImpB}. Again, there is no contribution to the momentum structures $K_{1i}K_{3j}$ or $\delta_{ij} \vec{K}_1 \cdot \vec{K}_3$.

\subsection{Correlator at separated points}
We can now understand the structure of the QFT correlator at low momentum. We have shown that at order $K^4$, with $K_{2u} = 0$, 
\be
G^{\rm QFT}_{uuij} = G^{\rm UV}_{uuij} + G^{\rm PC}_{uuij} + G^{\rm sep}_{uuij}\, ,
\ee
with $G^{\rm PC}_{uuij}$ given in \eqref{pcfinal}, and furthermore, in the same regime,
\be 
G^{\rm QFT}_{uuij} \approx G^{\rm IR}_{uuij}  + G^{\mbox{\scriptsize improvement}}_{uuij}\, ,
\ee
with the improvement term (if it is nonzero) given in \eqref{finalImpB}. Together these two relations imply 
\begin{align}\label{irsplit}
 G^{\rm IR}_{uuij} &\approx  G^{\rm UV}_{uuij}  + G^{\rm sep}_{uuij} + 
 \left( 
c_1'' K_{3i} K_{3j} 
+  c_2'' \delta_{ij} K_3^2  
+  c_3'' \delta_{ij}K_1^2  \right) K_{3u}^2 
\end{align}
where we have combined the PC terms and improvement terms. The approximation symbol `$\approx$' means that 
this result holds at low momentum at order $K^4$, and we have set $K_{2u} =0$.

As mentioned above, under a change of the renormalization scheme, it is possible for contributions to shift from the contact terms to the separated term and vice-versa. (For example, $K^2 \log (K^2/\mu^2)$ is shifted by an analytic term under rescaling $\mu$.) However, this does not affect the momentum tensor structure, so the coefficients of the momentum structures  $K_{1i}K_{3j}$ and $\delta_{ij} \vec{K}_1 \cdot \vec{K}_3$ have a completely unambiguous `separated' part that does not depend on the scheme. Thus \eqref{irsplit} shows unambiguously how these coefficients in the IR CFT are related to the UV CFT plus contributions from the QFT at separated points.

\section{Sum rule}\label{s:rg}

The CFT formula for the anomalies is now applied to the IR fixed point to derive sum rules for the change in the anomaly coefficient along the RG flow \cite{Hartman:2023qdn}. This strategy is quite general --- essentially any formula for CFT data can be applied to an IR fixed point and viewed as a sum rule for the RG flow, because the result in the IR must be obtained by integrating over all scales in the parent QFT. 

The starting point is the CFT formula for the anomalies given in \eqref{cftc} and \eqref{eq:SumRuleaPT}. Applying these results to the IR CFT, combined into a single formula for notational convenience, we have
\be\label{sr1}
c_{IR} + \left(\gamma - \tfrac{1}{2}\right) a_{IR} = \frac{1}{64}\int d^4 x_1 d^4 x_3\, \delta(u_1)u_3^2
\big(x_1^i x_3^j + \gamma \delta^{ij} \vec{x}_1 \cdot \vec{x}_3\big)
\langle \R[ \E_u; \Theta(x_1) T_{ij}(x_3)]\rangle_{\rm IR}\, .
\ee
The coefficient $\gamma$ is a placeholder introduced to keep track of the $(a,c)$ dependence on the two tensor structures in the integration kernel. In this expression, the integrand is a pure contact term, because the correlator is defined in the IR CFT.

Recall that this formula is designed to extract the coefficients of $k_{3u}^2k_{1i}k_{3j}$ and $k_{3u}^2\delta_{ij}\vec{k}_1 \cdot \vec{k}_3$ from the CFT correlator.   All other tensor structures in the 3-point function drop out of the integral. It was argued in section \ref{s:qft} that these terms in the IR CFT are given by the UV contact term plus the low-momentum limit of the QFT correlator at separated points. Thus, using \eqref{irsplit} in \eqref{sr1}, moving the UV terms to the other side of the equation, and flipping the overall sign, we find 
\be\label{srqft}
\Delta c + \left(\gamma - \tfrac{1}{2}\right)\Delta a 
 = 
- \frac{1}{64}\int d^4 x_1 d^4 x_3 \delta(u_1)u_3^2
\big(x_1^i x_3^j + \gamma \delta^{ij} \vec{x}_1 \cdot \vec{x}_3\big) \langle \R[ \E_u; \Theta(x_1) T_{ij}(x_3)]\rangle_{\rm QFT, sep}\, 
\ee
where
\begin{align}
\Delta a= a_{UV} - a_{IR} , \qquad\qquad  \Delta c = c_{UV} - c_{IR} \ . 
\end{align}
Equation \eqref{srqft} is exact, because it extracts just the $O(k^4)$ terms from the correlator. The integrand is supported only at separated points. In going from \eqref{sr1} to \eqref{srqft} we have resolved the delta function of the IR CFT into a sum of a UV delta function, plus a contribution at separated points. 
As discussed in section \ref{s:qft}, the split into a separated part and contact terms generally depends on the scheme. However, the momentum structure extracted by this integral kernel is scheme independent, so that despite some ambiguities in the integrand, the right-hand side in \eqref{srqft} is unambiguous.

At separated points, the retarded correlator is defined by the nested commutator \eqref{eq:Retardedat3pt}. Expanding out the commutators, most of the orderings vanish using $\E_u |0\rangle = 0$, so that at separated points the retarded correlator is simply \cite{Hartman:2023qdn}
\begin{align}
\begin{split}
 \langle \R[\E_u; \Theta(x_1) T_{ij}(x_3)]\rangle
 &= \theta(-v_1) \theta(-v_3) \left[  \langle \Theta(x_1) \E_u T_{ij}(x_3)\rangle + \langle T_{ij}(x_3)\E_u \Theta(x_1)\rangle \right] \\
 &= 2\theta(-v_1) \theta(-v_3) \mbox{Re}\,  \langle \Theta(x_1) \E_u T_{ij}(x_3)\rangle \ .
 \end{split}
\end{align}
The correlators on the right-hand side are Wightman functions, which do not have contact terms. Therefore, writing out the two anomalies separately and renaming $x_3 \to x_2$, we find
\begin{align}\label{csumrule}
\Delta c  &= -\frac{1}{32}\int_{v_1<0}\!d^4 x_1 \int_{v_2<0}\! d^4 x_2 \,\delta(u_1) u_2^2 \big(x_1^i x_2^j + \frac{1}{2} \delta^{ij} \vec{x}_1 \cdot \vec{x}_2\big)
\mbox{Re}\,
\langle \Theta(x_1) \E_u T_{ij}(x_2)\rangle \\
\label{asumrule}
\Delta a &= -\frac{1}{32}\int_{v_1<0}\!d^4 x_1 \int_{v_2<0}\! d^4 x_2\, \delta(u_1) u_2^2 \left( \delta^{ij} \vec{x}_1 \cdot \vec{x}_2\right)
\mbox{Re}\,
\langle \Theta(x_1) \E_u T_{ij}(x_2)\rangle \ .
\end{align}
This is our main result.
The sum rule for $\Delta c$ is new. The result for $\Delta a$ is equivalent to the sum rule derived in \cite{Hartman:2023qdn}, though not manifestly so. There, the sum rule for $\Delta a$ is written in terms of the full trace $\Theta$ rather than the transverse trace $\delta^{ij}T_{ij}$, which are related by $\Theta = -4 T_{uv} +\delta^{ij}T_{ij}$. Using the explicit expression for the CFT correlator in \eqref{fullthreept} it can be shown that the $T_{uv}$ term drops out of the integral, so in fact they agree.

It was also shown in \cite{Hartman:2023qdn} that the sum rule for $\Delta a$ can be written as an expectation value of the averaged null energy in a state created by a smeared insertion of $\Theta$ acting on vacuum, in order to prove positivity. This is clearly impossible for $\Delta c$, since it involves off-diagonal matrix elements of $\E_u$. This was to be expected, since it is  known that $\Delta c$ can be either sign.

These sum rules were derived from the retarded correlator, but one can just as easily start with time-ordered correlator. The resulting sum rules are equivalent to \eqref{csumrule}-\eqref{asumrule} under the contour rotation $\int_{-\infty}^{0} dv_1 \to - \int_0^{\infty} dv_1$.

In addition, using translation invariance of the 3-point function, the light-ray integral over $u$ can be shifted from $T_{uu}$ onto $\Theta$ to give the equivalent sum rules
\begin{align}\notag
\Delta c  &= -\frac{1}{32}\int_{v_1<0}\!d^4 x_1 \int_{v_2<0} \!d^4 x_2\, (u_1-u_2)^2  \big(x_1^i x_2^j + \frac{1}{2} \delta^{ij} \vec{x}_1 \cdot \vec{x}_2\big)
\mbox{Re}\,
\langle \Theta(x_1) T_{uu}(0) T_{ij}(x_2)\rangle \\
\label{asumruleBB}
\Delta a &= -\frac{1}{32}\int_{v_1<0}\!d^4 x_1 \int_{v_2<0} \!d^4 x_2\, (u_1-u_2)^2 \left( \delta^{ij} \vec{x}_1 \cdot \vec{x}_2\right)
\mbox{Re}\,
\langle \Theta(x_1) T_{uu}(0) T_{ij}(x_2)\rangle \ .
\end{align}

\section{Discussion}

The sum rules derived in this paper for $\Delta a$ and $\Delta c$ in four dimensions are in \eqref{csumrule}-\eqref{asumruleBB}. In this section, we summarize the other sum rules derived in various dimensions in our previous work \cite{Hartman:2023qdn,Hartman:2023ccw} for easy reference, and then briefly compare the results in four dimensions to the $S$-matrix approach in the recent work \cite{Karateev:2023mrb}.

\subsection{Collection of sum rules}

The RG sum rules derived in \cite{Hartman:2023qdn,Hartman:2023ccw} can be expressed in terms of the averaged null energy $\E_u$ or, by a simple shift of the integral, in terms of the null energy at the origin, $T_{uu}(0)$. In terms of the averaged null energy, the results in two and four dimensions, derived using the retarded correlators, are
\begin{align}
\Delta c_{\rm 2d} &=  -6\pi \int_{v_1<0}d^2 x_1 \int_{v_2<0}d^2 x_2 \, 
u_1^2 \delta(u_2)\langle \Theta(x_1)\E_u \Theta(x_2)\rangle\, \label{exactR}\\
\Delta a_{\rm 4d} &=  - \frac{1}{32} \int_{v_1<0} d^4 x_1 \int_{v_2<0} d^4 x_2 \, u_1^2 \delta(u_2) \vec{x}_1 \cdot \vec{x}_2 \langle \Theta(x_1)\E_u \Theta(x_2)\rangle \, ,
\end{align}
together with \eqref{csumrule}-\eqref{asumrule}. 
In each of these expressions one can replace $\int_{-\infty}^0 dv_1 \to - \int_{0}^\infty dv_1$. The sum rules in terms of the local null energy are
\begin{align}
\Delta c_{\rm 2d} &=  -6\pi \int_{v_1<0}d^2 x_1 \int_{v_2<0}d^2 x_2 \, 
(u_1-u_2)^2 \langle \Theta(x_1)T_{uu}(0) \Theta(x_2)\rangle \\
\Delta a_{\rm 4d} &= -\frac{1}{32}\int_{v_1<0} d^4 x_1 \int_{v_2<0} d^4 x_2 (u_1-u_2)^2 \vec{x}_1 \cdot \vec{x}_2 \langle \Theta(x_1) T_{uu}(0) \Theta(x_2)\rangle \ .\label{exactLocal}
\end{align}
Note that in the form we have written these sum rules, they are not manifestly positive. To make positivity manifest, they can also be written as expectation values of $\E_u$ in a state created by acting with $\Theta$ on the vacuum, smeared against a wavepacket. Thus the ANEC implies the $c$-theorem in two dimensions \cite{Hartman:2023ccw} and the $a$-theorem in four dimensions \cite{Hartman:2023qdn}.

\subsection{Comparison to $S$-matrix methods}
It was recently shown that the combination $\Delta c - \Delta a$ appears as the coefficient of a universal term in the dilaton-graviton scattering amplitude \cite{Karateev:2023mrb}.\footnote{See \cite{Gillioz:2018kwh,Karateev:2022jdb} for other approaches.} The amplitude is related to a complicated linear combination of 2-point, 3-point, and 4-point corelation functions of the stress tensor of the QFT; see \cite{Baume:2014rla} for the explicit relation in the case of dilaton-dilaton scattering, which has a universal term with coefficient $\Delta a$ \cite{Komargodski:2011vj}.
It would be very interesting to understand in detail how our sum rule, constructed from the 3-point function, agrees with the scattering amplitude. We will not attempt to explain this here, but we would like to compare to one of the intermediate results obtained in \cite{Karateev:2023mrb}, namely the dilaton-graviton-graviton vertex. This vertex should clearly be related to the correlation function that we considered above, $\lll \Theta(K_1) T_{\mu\nu}(K_2) T_{\alpha\beta}(K_3) \rr$. 

To compare, it is convenient to choose transverse-traceless polarizations for the two stress tensors, satisfying
\begin{align}
g_{\mu\nu}\epsilon_2^{\mu\nu} = g_{\mu\nu} \epsilon_3^{\mu\nu} = 0 , \qquad
K_{2\mu} \epsilon_2^{\mu\nu}= K_{3\mu} \epsilon_3^{\mu\nu} = 0 \ . 
\end{align}
Contracting these polarizations with the CFT 3-point function in \eqref{fullthreept} gives 
\begin{align}\label{ttcorr}
&\epsilon_2^{\mu\nu}\epsilon_3^{\alpha\beta}\lll \Theta(K_1)T_{\mu\nu}(K_2)T_{\alpha\beta}(K_3)\rr\nonumber\\
&\quad  =  (\epsilon_{2\mu\nu} \epsilon_3^{\mu\nu} )f_1 + (K_{2\alpha}\epsilon_3^{\alpha\beta} K_{2\beta})(K_{3\mu} \epsilon_2^{\mu\nu}  K_{3\nu}) f_2  + (K_{2}^\alpha \epsilon_{3\alpha\beta} \epsilon_2^{\beta\rho}  K_{3\rho} )  f_3\, ,
\end{align}
where the three coefficient functions are 
\begin{align}
f_1 &=\left(8(c-a)-20b_1\right)(K_2\cdot K_3)^2+ 4 (2a-c-2b_1)K_2^2K_3^2-4b_1(K_2^4+K_3^4) \notag\\
&\quad-18b_1(K_2\cdot K_3)\left(K_2^2+K_3^2\right)+4b_2\Lambda^2\left(K_2^2+K_3^2\right)+10b_2(K_2\cdot K_3)\Lambda^2+2b_3\Lambda^4\nonumber\\
f_2 &= 8(c-a)\,\label{eq:f2}\\
f_3 &=8\left(2(a-c)+b_1\right) (K_2\cdot K_3)+4b_1(K_2^2+K_3^2)-4b_2\Lambda^2\, . \notag
\end{align}
The full correlator  \eqref{fullthreept} can be reconstructed from its transverse-traceless part \eqref{ttcorr} by writing it in terms of projectors and using the Ward identity \eqref{eq:WI3} and the 2-point function \eqref{eq:Euclidean4dTT2k}. 

The dependence on $a$ and $c$ in \eqref{ttcorr} agrees perfectly with the vertex obtained in \cite{Karateev:2023mrb}. In the $S$-matrix approach, this vertex is used as one of several building blocks to calculate the dilaton-graviton amplitude. It is not transparent exactly how (or whether) the combination $(c-a)$ appearing in the amplitude is related to the coefficient function $f_2 \sim c-a$ in \eqref{eq:f2}. But we will now argue that our null energy sum rule for $(c-a)$ that is simply extracting $f_2$ from this vertex. The comparison is not entirely trivial because it is effectively written in a different gauge.

Suppose we expand the full 3-point function $\lll \Theta T_{\mu\nu} T_{\alpha\beta}\rr$ in all possible 4-index monomial tensors built from the metric $g_{\alpha\beta}$ and the momenta $K_{2\alpha}$ and $ K_{3\alpha}$. Clearly $f_2$ in \eqref{ttcorr} is the coefficient of the tensor $K_{2\alpha} K_{2\beta} K_{3\mu} K_{3\nu}$, since this is the only monomial that can contribute to this tensor structure. So we must show that our sum rule is also extracting the coefficient of this tensor. 

Our sum rule for the combination $(c-a)$ is obtained by setting $\gamma = -\tfrac{1}{2}$ in \eqref{srqft}, 
\be \label{diffsum}
\Delta c - \Delta a = 
 -\frac{1}{32}\int_{v_1<0}d^4 x_1 \int_{v_3<0} d^4 x_3\, \delta(u_1) u_3^2 \big(x_1^i x_3^j - \frac{1}{2} \delta^{ij} \vec{x}_1 \cdot \vec{x}_3\big)
\mbox{Re}\,
\langle \Theta(x_1) \E_u T_{ij}(x_3)\rangle\, .
\ee
Note that the kernel is traceless in the two-dimensional transverse space. 
It is illuminating to consider another way to obtain this combination directly, without first calculating $\Delta a $ and $\Delta c$ separately. Let $w_{ij}$ be a traceless 2x2 polarization tensor restricted to the directions orthogonal to the $uv$-plane. For $K_{2u}=0$, the CFT correlator  \eqref{mainEuc} contracted into this polarization, at order $K^4$, is 
\begin{align}\label{wpol} 
 w^{ij} \lll \Theta(K_1) T_{uu}(K_2) T_{ij}(K_3)\rr 
&= 8 (c-a) w^{ij} K_{2i}K_{2j} K_{3u}^2 - 4b_1 w^{ij} K_{3i} K_{3j} K_{3u}^2   \ . 
\end{align}
The kernel in \eqref{diffsum} extracts the coefficient of the first term (times $\frac{1}{8}$), because the second term has no $K_{1i}K_{3j}$. Furthermore, the coefficient of this term is clearly equal to the coefficient of $K_{2\alpha} K_{2\beta} K_{3\mu} K_{3\nu}$ in the expansion of the full correlator in monomial 4-index tensors built from the metric and $K_{2\alpha}$, $K_{3\alpha}$, the same as $f_2$. Therefore the sum rule \eqref{diffsum} calculates $\frac{1}{8} \Delta f _2$.

This makes a connection to the effective action formalism of \cite{Karateev:2023mrb} but does not answer the deeper, more general question of how light-ray sum rules are related to scattering amplitudes. It would be interesting to explore this further. It would also be worthwhile to study examples with spontaneous breaking of conformal symmetry --- it is not entirely clear how to match $\Delta c$ in such cases \cite{Schwimmer:2010za,Niarchos:2020nxk,Schwimmer:2023nzk,Karateev:2023mrb}.

\ \\

\noindent \textbf{Acknowledgments}
\noindent  We thank Austin Joyce, Denis Karateev, Zohar Komargodski, Manuel Loparco, David Meltzer, Jo\~{a}o Penedones, Riccardo Rattazzi, Biswajit Sahoo, and John Stout for discussions. TH is supported by NSF grant PHY-2014071. GM is supported by the Simons Foundation grant 488649 (Simons
Collaboration on the Nonperturbative Bootstrap) and the Swiss National
Science Foundation through the project 200020\_197160 and through the
National Centre of Competence in Research SwissMAP. This research was also supported in part by grant NSF PHY-1748958 to the Kavli Institute for Theoretical Physics (KITP).

\appendix

\section{Ambiguities in the QFT correlator}

In section \ref{s:qft} in the main text, we discussed how the QFT correlator at low momentum is related to the correlator of the IR CFT. The relation is complicated by improvement terms to the stress tensor in the IR, and semi-local or partial contact terms with a single delta function in position space. In this appendix we enumerate the various ambiguities that appear in both cases in order to derive the result used in the main text, which is that for $K_{2u} = 0$, expanding to $O(K^4)$, neither of these issues affects the momentum structures $K_{1i} K_{3i}$ or $\delta_{ij} \vec{K}_1 \cdot \vec{K}_3$ that were used to extract both anomaly coefficients. This was necessary to obtain a meaningful sum rule in section \ref{s:rg}. 

\subsection{Improvement terms}\label{app:improvement}

According to the discussion around \eqref{ishift}, the low-momentum limit of the QFT correlator is related to the IR CFT correlator by
\be
\lll \Theta(K_1) T_{uu}(K_2) T_{ij}(K_3) \rr_{\rm QFT}
\approx 
\lll \Theta(K_1) T_{uu}(K_2) T_{ij}(K_3) \rr_{\rm IR} + (\mbox{improvements})\, ,
\ee
where the improvements come from possible shifts in \eqref{ishift}.
For $K_{2u} = 0$, the improvement from $T_{uu}(K_2)$ vanishes. Shifting the other two stress tensors gives all the possible contributions from improvement terms
\begin{align}\label{imp1}
(\mbox{improvements}) &= 
K_1^2 \lll \Phi(K_1) T_{uu}(K_2) T_{ij}(K_3)\rr_{\rm IR}
+ \Pi_{3ij} \lll \Theta(K_1) T_{uu}(K_2) \Phi(K_3)\rr_{\rm IR}\notag \\
&\qquad 
 + K_1^2 \Pi_{3ij} \lll \Phi(K_1) T_{uu}(K_2)  \Phi(K_3) \rr_{\rm IR}\, ,
\end{align}
with $\Pi_{A \mu\nu} = K_{A\mu} K_{A\nu} - g_{\mu\nu} K_A^2$, and $A=\lbrace 1,\,2,\, 3\rbrace $ labels the three momenta. We have set $K_{2u} = 0$ and suppressed the numerical coefficients in this expression. The correlator in the last term in \eqref{imp1} at $O(K^0)$ must be proportional to $g_{uu}$, so it vanishes. Moreover, the first term already has two powers of momenta stripped out such that the three-point function must contribute at order $O(K^2)$, and the only such invariant with free indices $uuij$ is $K_{3u}^2 \delta_{ij}$. Similarly, the only nonzero tensor structure in the correlator appearing in the second term at $O(K^2)$ is $K_{3u}^2$. We thus conclude
\be \label{appimp}
(\mbox{improvements}) =c_1 K_1^2 K_{3u}^2 \delta_{ij} + c_2 K_{3u}^2 (K_{3i} K_{3j} - \delta_{ij} K_3^2) \ ,
\ee
for some constants $c_{1,2}$.  Therefore the QFT correlator and the IR CFT correlator are related as stated in \eqref{finalImpA}-\eqref{finalImpB}.

\subsection{Semi-local terms}\label{app:semilocal}

Here we will analyze the possible delta-function contributions to the OPE, as in \eqref{eq:OPE1} and \eqref{eq:OPE2}, satisfying the unitarity bound of the UV CFT, in order to derive \eqref{pcfinal}. 
We start with the contributions in \eqref{eq:OPE1}, which we rewrite here specialized to null indices on the stress tensor:
\be 
T_{uu}(x_2) \Theta(x_1) \sim \O^{\sigma_1\dots \sigma_{\ell}}(x_2) P_{uu\sigma_1\dots \sigma_{\ell} }(\p) \delta^{(4)}(x_2-x_1) \ .
\ee
In the 3-point function, this contributes a term
\be 
\langle \Theta(x_1) T_{uu}(x_2) T_{ij}(x_3) \rangle  \supset \langle \O^{\sigma_1\dots \sigma_\ell}(x_2) T_{ij}(x_3)\rangle P_{uu \sigma_1\dots \sigma_\ell}(\p) \delta^{(4)}(x_2-x_1) \ . 
\ee
Transforming to momentum space, this becomes
\be\label{type1PC}
G_{uuij}(K_1,K_3) \supset H^{\sigma_1\dots \sigma_\ell}_{ij}(K_3) P_{uu\sigma_1\dots \sigma_{\ell}}(iK_1) \ ,
\ee
where $H^{\sigma_1\dots \sigma_\ell}_{ij}(K)= \lll \O^{\sigma_1\dots \sigma_{\ell}}(K)T_{ij}(-K)\rr$. We now expand at low momentum and look for terms with exactly four powers of momentum. At low momentum, we can evaluate the 2-point function $H^{\sigma_1\dots \sigma_\ell}_{ij}$ at the IR fixed point. Thus $H$ has integer dimension $\Delta_\O$, where this is the scaling dimension of $\O$ in the IR. Therefore the tensor $P$, which is built from $K_1$ and the metric, has dimension 
\be \label{req1}
 \dim P = 4 - \Delta_\O \ . 
\ee
The unitarity bound in the IR for the operator $\O$ (which is not required to be primary) is 
\be \label{req2}
\Delta_\O \geq \ell+1 \ ,
\ee 
so $\dim P \leq 3 - \ell$. 
Since $P$ is a tensor built from $g_{\mu\nu}$ and non-negative powers of $K_{1}$, if $P$ is an even (odd) power of $K$ then it has an even (odd) number of total indices. That is,
\be \label{req3}
\dim P = \ell \mod 2 \ .
\ee
There is a finite list of dimensions and spins satisfying the requirements \eqref{req1}-\eqref{req3}:
\be \label{allowedDims}
(\ell, \Delta_{\O}, \dim P) = (0,4,0),\  (0,2,2), \  (1,3,1), \mbox{\ or\ } (2,4,0) \ . 
\ee
These contribute terms
\begin{align}\label{allpcs1}
\lll \Theta(K_1) &T_{uu}(K_2) T_{ij}(K_3)\rr \supset \notag\\
&\lll \O_{[4]}(K_3) T_{ij}(-K_3) \rr P_{uu}^{[0]} (iK_1) 
+ \lll \O_{[2]}(K_3) T_{ij}(-K_3) \rr P_{uu}^{[2]}(iK_1) \\
&+ \lll \O_{[3]}^{\sigma_1} (K_3) T_{ij}(-K_3)\rr P_{uu\sigma_1}^{[1]}(iK_1) 
+ \lll \O_{[4]}^{\sigma_1\sigma_2}T_{ij}(-K_3)\rr P_{uu \sigma_1 \sigma_2}^{[0]}(iK_1) \notag \ ,
\end{align}
where subscripts as in $\O_{[\Delta]}$ denote the IR scaling dimension of the operator and superscripts as in $P^{[n]}$ denote polynomials in $K$ of degree $n$. 
The first case $(0,4,0)$ is excluded because then $P_{uu}^{[0]} = g_{uu} = 0$. 
For $(\ell,\Delta_{\O}) = (0,2)$, there is only one structure for the contact term, $P_{uu}^{[2]}= K_{1u}^2$, and $H_{ij}(K_3) 
= \lll \O(K_3) T_{ij}(-K_3)\rr$ expanded at low momentum has two possible structures. Using \eqref{type1PC} this leads to terms
\be \label{type1spin0}
G_{uuij} \supset  c'_1 K_{3i} K_{3j} K_{1u}^2 + c'_2 K_3^2 \delta_{ij} K_{1u}^2 \ . 
\ee
For $(\ell, \Delta_{\O})=(1,3)$, we have $P_{uu\sigma_1}^{[1]} = K_{1u} g_{u\sigma_1}$,  and
the contribution to the correlator \eqref{type1PC} is
\be
G_{uuij} \supset  H_{uij}(K_3) K_{1u} \ . 
\ee
Using $K_{3u} = - K_{1u}$, this produces terms of exactly the same form as \eqref{type1spin0}, which we can therefore absorb into the coefficients $c'_1$ and $c'_2$.  Finally, consider $(\ell,\Delta_{\O})=(2,4)$. The tensor $P$ must be dimensionless, so $P_{uu\sigma_1\sigma_2}^{[0]} = g_{u\sigma_1}g_{u\sigma_2}$, and the correlator has a term
\be
G_{uuij} \supset H_{uuij}(K_3) \ .
\ee
Again this has the same form as \eqref{type1spin0}, using $K_{1u}=-K_{3u}$. Therefore \eqref{type1spin0} is the only contribution to the correlator from the semilocal terms in $\{ \Theta T_{uu} \}$.

We now consider the second type of partial contact term $\{T_{\mu\nu}T_{\alpha\beta}\}$ given in \eqref{eq:OPE2}, that is,
\be
T_{uu}(x_2) T_{ij}(x_3) \sim \O^{\sigma_1\dots \sigma_\ell}(x_2) U_{uuij\sigma_1\dots\sigma_\ell}(\p) \delta^{(4)}(x_2-x_3) \ . 
\ee
Following the same logic, the contribution of this term to the momentum space 3-point function is
\be
G_{uuij} \supset L^{\sigma_1\dots \sigma_\ell}(K_1) U_{uuij\sigma_1\dots \sigma_\ell}(iK_3)  \ ,
\ee
where $L^{\sigma_1\dots \sigma_\ell} = \lll \Theta \O^{\sigma_1\dots \sigma_{\ell}} \rr$. 
The list of operator spins and dimensions $(\ell,\Delta_{\O},\dim U)$ that can potentially contribute at $O(K^4)$ is the same as listed in \eqref{allowedDims}, so the four possible terms are
\begin{align}\label{allpcs2}
\lll \Theta(K_1) &T_{uu}(K_2) T_{ij}(K_3)\rr \supset \notag\\
& \lll \Theta(K_1)\O_{[4]}(-K_1)\rr U_{uuij}^{[0]}(iK_3)
+ \lll \Theta(K_1)\O_{[2]}(-K_1)\rr U_{uuij}^{[2]}(iK_3)  \\
&+ \lll \Theta(K_1) \O_{[3]}^{\sigma_1}(-K_1) \rr U_{uuij\sigma_1}^{[1]}(iK_3) 
+ \lll \Theta(K_1) \O_{[4]}^{\sigma_1\sigma_2}(-K_1)\rr U_{uuij\sigma_1\sigma_2}^{[0]}(iK_3) \ .
\notag
\end{align}
 For $(\ell, \Delta_{\O}) = (0,4)$, we have $U_{uuij}^{[0]} = g_{uu} \delta_{ij}$ or  $U_{uuij}^{[0]} =  g_{ui}g_{uj} $, both of which vanish. For $(\ell,\Delta_{\O}) = (0,2)$, we have  $U_{uuij}^{[2]} = K_{3u}^2 \delta_{ij}$, and the corresponding contribution to the correlator is
\be\label{type2spin0}
G_{uuij} \supset c'_3 K_1^2 K_{3u}^2 \delta_{ij} \ . 
\ee
For $(\ell,\Delta_{\O})=(1,3)$, we have $\dim U = 1$, and the only such tensor is  $U_{uuij\sigma_1}^{[1]} = K_{3u} g_{u\sigma_1} \delta_{ij}$. In the correlator this produces a term of the same form as \eqref{type2spin0}. The last contribution to consider is $(\ell,\Delta_{\O})=(2,4)$, which requires $\dim U = 0$, so  $U_{uuij\sigma_1\sigma_2}^{[0]} = g_{u\sigma_1} g_{u\sigma_2} \delta_{ij}$. This also leads to the same term as \eqref{type2spin0}.

Putting it all together, we have shown in this appendix that the only partial contact terms that can appear in the ANE correlator are those in \eqref{type1spin0} and \eqref{type2spin0}. Using $K_{1u} = -K_{3u}$ the general contribution is \eqref{pcfinal} in the main text.

Note that we have not imposed the Ward identity on the PC terms, because it was not necessary. Imposing the Ward identity would further restrict the coefficients such that the allowed PC terms are identical to the improvement terms in \eqref{appimp}.

\addcontentsline{toc}{section}{References}
\bibliographystyle{utphys}
{\small
\bibliography{deltaC}
}

\end{document}